\documentstyle[11pt,inaoe_proc,psfig2,epsfig,twoside]{article} 
\def\apj{ApJ}
\def\B{{\bf B}}
\def\Bo{B_{\rm o}}
\def\BP{Ballesteros-Paredes}
\def\e{\hat {\bf e}}
\def\M{{\cal M}}
\def\s{{\bf S}}
\def\T{{\bf T}}
\def\u{{\bf u}}
\def\va{v_{\rm A}}
\def\VS{V\'azquez-Semadeni}
\def\x{{\bf x}}

\def\ltsima{$\; \buildrel < \over \sim \;$}    
\def\simlt{\lower.5ex\hbox{\ltsima}}           
\def\gtsima{$\; \buildrel > \over \sim \;$}    
\def\simgt{\lower.5ex\hbox{\gtsima}}           

\begin{document}

\title{Energy Budget and the Virial Theorem in Interstellar Clouds}  

\author{Enrique V\'azquez-Semadeni} 
\affil{ Instituto de Astronom\'ia, UNAM\\
Apartado Postal 70-264, M\'exico D.F. 04510, MEXICO  }  

\begin{abstract}
The Virial Thoerem (VT) is a mathematical expression obtained from the
equation of motion for a fluid, which describes the energy budget of
particular regions within the flow. This course
reviews the basic theory leading to the VT, discusses
its applicability and limitations, and then summarizes observational
results concerning the physical and statistical properties of
interstellar clouds which are normally understood in terms of the
VT, in particular the so-called ``Larson's Relations''. In all cases,
the standard notions as well as relevant counterpoint are presented. In
particular, the difficulties arising when the medium is
fully turbulent are discussed.
\end{abstract}

\keywords{Hydrodynamics -- ISM: clouds -- ISM: kinematics and dynamics
  -- ISM: magnetic fields -- turbulence} 

To appear in ``Millimetric and Sub-Millimetric Astronomy. INAOE 1996 Summer
School''.

\section{ Introduction }
\markboth{Enrique V\'azquez-Semadeni}{Virial Theorem in Interstellar Clouds}

The interstellar medium is an extremely complex mixture of gas, dust,
and cosmic rays, threaded by a ubiquitous magnetic field, and stirred by a
wide variety of energy sources such as supernovae, expanding HII
regions and bipolar outflows. 
Progressively higher-density regions in this medium
constitute what we normally call intercloud medium,
cloud complexes, diffuse clouds,
molecular clouds, clumps and cores, respectively, although the
boundaries between these classifications are fuzzy (a natural
consequence of drawing arbitrary classification boundaries in a
continuum). Table \ref{ism properties} gives the ranges of densities
and temperatures for the above structures (adapted from Jura 1987).

\begin{table}\label{ism properties}
\caption[ ]{Structures in the interstellar medium. \\
(Adapted from Jura 1987.)}
\begin{flushleft}

\begin{tabular}{llll}
\hline
\noalign{\smallskip}
Name & $n({\rm cm}^{-3})$ & $T$(K) & Filling factor \\
\noalign{\smallskip}
\hline
\noalign{\smallskip}
Hot intercloud & $3\times 10^{-3}$ & $10^6$ & $0.1-0.7$? \\
gas & & & \\
Warm gas & $0.1$ & $10^3-10^4$ & $0.4$  \\
& & & \\
& & & \\
Diffuse clouds & $< 10$ & $50-100$ & \\
& & & \\
& & & \\
Cirrus clouds & $10-10^3$ & $10-100$ & \\
& & & \\
Dark clouds & $>10^3$ & $10$ & $0.01$? \\
& & & \\
GMCs & $>10^3$ & $15-40$ & \\
H~II regions & $>10$ & $10^4$ & \\
& & & \\
SNR & $>1$ & $10^4-10^7$ & \\
\noalign{\smallskip}
\hline
\end{tabular}

\end{flushleft}
\end{table}
The gaseous component of this medium
should be reasonably well described as a compressible,
self-gravitating fluid  (e.g, Shu 1992, Ch. 1)
obeying the magnetohydrodynamic (MHD) equations, 
since even very low ionization fractions allow coupling of
the motions to the magnetic field (Mestel \& Spitzer 1956; see also
Shu 1992, Ch.\ 27). From one of these equations, a
very important result called the Virial Theorem (VT) can be obtained,
which describes the energy balance (or ``budget'') of particular
regions within the medium. In this course, we first review the
derivation of the VT (\S\ \ref{derivation}), 
stressing in particular the physical meaning of
each term, including the often-neglected surface terms,
and the validity of frequently used
simplifying assumptions (e.g., Shu 1992). In \S\ \ref{applications} we
then discuss various applications of the
Virial Theorem, both to idealized cases and to real interstellar cloud
properties, pointing out existing incompletenesses. Finally, in
\S\ \ref{conclusions} contains the conclusions. Throughout this
course, we shall ignore the flux-loss effects of ambipolar
diffusion (e.g, Shu 1992).

\section{The MHD Equations and the Virial Theorem}\label{derivation}

\subsection{The MHD equations}

In order for the gaseous component of the ISM to be adequately
described by a continuum approximation, it is necessary that the mean
free paths of the gas molecules be much smaller than the
characteristic length scale of the systems under consideration (e.g.,
Currie 1974). This is generally satisfied in most kinds of
interstellar structures, since the characteristic length scales are
normally at least some $10^5$ times larger than the mean free paths
(e.g., Shu 1992). The use of the MHD equations is thus amply
justified. These equations, describing the evolution of a magnetized,
compressible gas without dissipative terms, 
are (e.g., Cowling 1976; Spitzer 1978; Shu 1992; Shore 1992):

\noindent
i) {\it The Continuity Equation}, expressing mass conservation:
\begin{equation} \label{continuity}
{\partial\rho\over\partial t} + {\nabla}\cdot (\rho\u) = 0;
\end{equation}
\noindent
ii) {\it The Momentum Balance Equation}
\begin{equation}\label{momentum}
{\rho \partial\u\over\partial t} + \rho \u\cdot\nabla\u =
-{\nabla P} - \rho \nabla \varphi + {1 \over 4 \pi} \bigl(\nabla
\times {\bf B}\bigr) \times {\bf B} - 2 \rho \Omega \times \u;
\end{equation}
\noindent
iii) {\it The Internal Energy Balance Equation} \label{eint}
\begin{equation}\label{internal energy}
{\partial e\over\partial t} + \u\cdot\nabla e = -(\gamma -1)e\nabla\cdot \u 
+ \Gamma - \rho \Lambda;
\end{equation}
\noindent
iv) {\it The Magnetic Field Equation (flux freezing)}
\begin{equation}\label{magnetic}
{\partial {\bf B}\over\partial t} = \nabla \times (\u \times {\bf B});
\end{equation}
\noindent
v) {\it Poisson's Equation}
\begin{equation}\label{poisson}
\nabla^2 \varphi=4 \pi G \rho.
\end{equation}
Additionally, we assume an ideal-gas equation of state
$P=(\gamma-1)\rho e$.

In these equations, $\rho$ is the mass density of the gas, \u\ is the
fluid velocity, $P$ is the thermal pressure, $e$ is the specific
internal energy, \B\ is the magnetic field, and $\varphi$ is the
gravitational potential. Additionally, $\Gamma$ and $\Lambda$
symbolically represent any heating and cooling sources present in the
medium, and are normally functions of the density and temperature in
the MHD description (neglecting radiative transfer and chemical composition).

Let us now briefly discuss the meaning of the main terms in these
equations. In eq.\ (\ref{momentum}), which is Newton's Second Law in
per-unit-volume form, the second term on the left hand
side (LHS) is the so-called non-linear or {\it advective} term, and
expresses the momentum transport by the velocity field itself. This is
the term that describes the effects of turbulence in the medium. The
first term in the right hand side (RHS) gives the force (per unit
volume) exerted by the
thermal pressure gradient. The second term is the
self-gravitational force, the third term is the Lorentz force, 
and the last term is the Coriolis force, due to
Galactic rotation with an angular velocity $\Omega$. 

In eq.\ (\ref{internal energy}), the second term on the LHS expresses
internal energy transport by the velocity field (convection). The
first term on the RHS is the $PdV$ work. In eq.\ (\ref{magnetic}), the RHS
is the {\it flux freezing} condition, implying that magnetic field
lines are dragged along with the matter. Finally, eq.\ (\ref{poisson})
expresses that the source of the gravitational potential is the mass
density.

Important physics left out from this description are, as mentioned
above, the radiative transfer and the chemical and ionization fraction
evolution. These can be crudely incorporated in the equations above by
means of model terms, such as $\Gamma$ and $\Lambda$ in eq.\
(\ref{internal energy}), or by making the equation of state vary in
different regimes. This allows an approximate description of the
effects of these processes on the dynamics, although it cannot allow
any predictions on those processes themselves. Finally, note that the
MHD equations hold in the so-called ``MHD approximation'' which
assumes an conductivity of the medium.

\subsection{The Virial Theorem}\label{VT}

The scalar VT is obtained by dotting the momentum equation (eq.\ \ref{momentum})
with the position
vector \x\ and integrating over some volume $V$. Traditionally, the volume
$V$ is taken in a {\it Lagrangian} frame of reference moving with the
fluid (e.g., Spitzer 1978; Shu 1992). An {\it Eulerian} (i.e,
in a fixed frame of reference) version of the VT has recently been
derived by McKee \& Zweibel (1992, hereafter MZ92). This may be a more
convenient form for computation in practice (e.g., MZ92;
\BP\ \& \VS\ 1997), but here we discuss the
Lagrangian version for consistency with most treatments. Neglecting
the Coriolis force, which is negligible at molecular cloud scales, one
obtains
\begin{equation}
\int_V \rho \x \cdot {d\u \over dt} dV = - \int_V \x \cdot \nabla P dV - 
\int_V \rho \x \cdot \nabla \varphi dV + {1 \over 4 \pi} \int_V \x \cdot
(\nabla \times \B) \times \B dV,
\end {equation}
where $d/dt=\partial/\partial t+\u \cdot \nabla$ is the total or
Lagrangian derivative operator. The LHS can be rewritten as
\begin{equation}\label{lhs}
\int_V \rho \x \cdot {d\u \over dt} dV = {1 \over 2} \int {d^2 x^2 \over
  dt^2} dm - \int_V \rho u^2 dV \equiv {1 \over 2} {d^2I \over dt^2} -
\int_V \rho u^2 dV,
\end{equation}
where we have defined $dm = \rho dV$, and have used Reynolds'
Transport Theorem 
\begin{equation}
{d \over dt} \int_V \alpha dV = \int_V \Bigl[{\partial \alpha \over \partial
  t} + \nabla \cdot (\alpha \u)\Bigr] dV
\end{equation}
to show that ${d\over dt} \int \alpha dm = \int 
{d\alpha\over dt}  dm$. Also, the
second equality in eq.\ (\ref{lhs}) defines the moment of inertia $I$. 

For the RHS,
consider the following. First, according to Newton's Law of
Gravitation (or from the solution of Poisson's equation) 
the gravitational term can be written as 
\begin{equation}
W \equiv - \int_V \rho(\x) \x \cdot \nabla \varphi dV= 
- G \int_V \int_{\rm all\ space}
{\rho(\x) \rho(\x ') \x \cdot (\x - \x ') \over |\x - \x '|^3} dV' dV.
\end{equation}
Noting that the integrand is nearly symmetric with respect to the primed and
unprimed variables, we can replace \x\ by $(\x - \x')/2$ and write
\begin{equation}
W = - {1 \over 2} G \int_V \int_{\rm all\ space} {\rho(\x) \rho(\x')
  \over |\x - \x'|} dV dV'. \label{def.W}
\end{equation}
Thus, $W$ is the gravitational energy of the mass distribution within
volume $V$, given the mass distribution of the whole
space.\footnote{Interestingly, this is not true in two dimensions (2D)
  (\BP\ \& \VS, in preparation), due to the different spatial
  variation of the gravitational potential, analogously to the
  well-known situation for electromagnetic fields. This is
  unfortunate, since many
  numerical simulations are performed in 2D due to computational limitations.}
Note that in Shu (1992), both integrals are evaluated over volume
$V$. This approximation is valid only if the mass
outside volume $V$ can be neglected (see Spitzer 1978), a fact which
is often overlooked.

Secondly, the remaining thermal pressure and magnetic terms can be
dealt with using Gauss' divergence theorem, and the facts that $\nabla
\cdot \x = 3$ and that (in tensor notation)
$\partial x_i/\partial x_j = \delta_{ij}$. Additionally, for the
magnetic term, it is most convenient to work with Maxwell's magnetic
stress tensor $\T \equiv T_{ij} \equiv B_i B_j/4 \pi - |\B|^2 \delta_{ij}/8
\pi$, such that the flux freezing term satisfies $(\nabla \times \B)
\times \B)/4 \pi = \nabla \cdot T_{ij}$. Thus, the VT finally reads
\begin{eqnarray}
\lefteqn{{1 \over 2} {d^2I \over dt^2} =  \int \rho u^2 dV + \Bigl(3 \int P dV -
\oint_S P \x \cdot d\s \Bigr) }\nonumber \\
& &+ \Bigl({1 \over 8 \pi} \int B^2 dV +
\oint_S \x \cdot \T \cdot d\s\Bigr) \nonumber \\
& &- {1 \over 2} G \int_V \int_{\rm
  all\ space} {\rho(\x) \rho(\x') \over |\x - \x'|} dV dV'  \nonumber \\
& &\equiv 2K + \Bigl(2U - \oint_S P \x \cdot d\s \Bigr) + \Bigl(\M + 
\oint_S \x \cdot \T \cdot d\s\Bigr) + W,\label{VTeq}
\end{eqnarray}
where $S$ is the surface enclosing $V$.

Several comments and warnings
are in order. First, note that the LHS is the second
time derivative of the moment of inertia of the cloud. The cloud is
said to be in {\it virial equilibrium} if the RHS is zero. However,
note that in particular this includes the case of a constant rate
of change in the cloud's moment of inertia. Thus, virial equilibrium
does not necessarily imply a stationary or static cloud configuration,
although the expression is almost always interpreted to mean so.

Second, note that both the thermal pressure and the magnetic terms
contain a ``volumetric'' and a ``surface'' contribution. In fact, if
either the thermal pressure or the magnetic field are uniform, then
the corresponding surface and volumetric terms cancel each
other, indicating balance between internal and external stresses.
Conversely, the
surface terms can only be ignored if the corresponding fields can be
neglected at the boundary surface. However, it is frequently
encountered in the literature that virial balance is considered only
among the volumetric terms.

Third, the thermal pressure surface term describes the
``confining'' effect of the external pressure on the cloud (volume
$V$). A similar comment applies to the magnetic surface term, although
in this case this term is not exclusively confining, as it includes
the effects of both the external magnetic pressure as well as the
magnetic tension along field lines. Moreover, 
in the Eulerian version of the theorem, an analogous surface term of
the form $\int \x \cdot \rho \u \u \cdot d\s$ for
the macroscopic velocity field also appears (MZ92), 
where $\rho \u \u$ is a momentum transfer tensor reminiscent of the
Reynolds' stress tensor (see, e.g., Landau \& Lifshitz 1987, \S\ 42;
Lesieur 1990, Ch.\ VI). However, due to the strongly fluctuating 
nature of the turbulent velocity field, which may contain large-scale
motions (comparable to the cloud's size), this term can hardly be considered as
producing a confining effect; instead, it produces
transport of momentum across the cloud's boundary and  a
redistribution of the mass within the cloud, thus contributing to the
change in its moment of inertia. Neglect of this type of exchange
across an Eulerian boundary, or of the high mobility of a Lagrangian
boundary, led MZ92 to the conclusion that the intercloud
velocity component perpendicular to the surface of the cloud must
be small at small distances from the boundary surface S, a result
which appears questionable in the light of numerical simulations of
turbulence in the ISM (Passot et al. 1988; L\'eorat et al 1990;
\VS\ et al. 1995; Passot et al. 1995), which
suggest an extremely dynamical scenario.

The last point leads to a practical warning. The choice of the
volume $V$ and its boundary $S$ in the actual ISM or in numerically
simulated flows constitutes a difficult task as soon as
high enough resolution is available. The VT has traditionally been
applied to idealized spheroidal, static clouds either in magnetohydrodynamic
equilibrium (possibly assisted by external pressure confinement; 
e.g., Strittmatter 1966; Mouschovias \& Spitzer 1976; see also Shu 1992),
equilibrium between internal ({\it micro})turbulence and self-gravity
(Chandrasekhar 1951; Bonazzola et al.\ 1987; \VS\ \& Gazol 1995), or
combinations thereof (e.g, Myers \& Goodman 1988a,b; Mouschovias \&
Psaltis 1995). In these cases, the clouds have well defined
boundaries. However, if the clouds are highly dynamical, even
possibly with fractal structure (Scalo 1990; Falgarone et al.\ 1991),
and they are observed with high enough resolutions (e.g., Bally et
al.\ 1987; see also the numerical results of \BP\ \& \VS\ 1997), then
important difficulties arise. In a Lagrangian description, the cloud
boundaries (the surfaces $S$) move in an extremely complex way,
getting stretched and distorted. In an Eulerian description, a surface
$S$ defined at a particular time will have an extremely amorphous shape,
possibly being very elongated or having ``tentacle''-like
protrusions. Thus, at subsequent times, mass will have entered or left
the contained volume $V$, rendering the identity of the cloud dubious
(\BP\ \& \VS\ 1997). In this case, it is
likely that all the terms in the RHS of the VT have comparable
importance, particularly the surface terms. 
With this caveat in mind, we now proceed to review some of
the best known applications of the VT to interstellar clouds.

\section{Applications}
\label{applications}

\subsection{Ideal cases}

\subsubsection{Pressure confinement.}

Taking $\B = \u =0$ and neglecting self-gravity in eq.\ (\ref{VTeq}), one gets
the condition of virial equilibrium, $d^2I/dt^2 = 0 \Rightarrow 
2U = \oint_S P\x \cdot d\s$, indicating balance between internal and
external pressure. This is a good example of a case in which volume
$V$ can be chosen in a completely arbitrary way, since this is the
case, for example, of any  parcel of gas in the air in the room you
are in.

\subsubsection{Hydrostatic equilibrium and turbulent support.}

Assuming $\B = \u =0$ and a vanishing thermal pressure outside volume
$V$, but keeping self-gravity, we obtain $d^2I/dt^2=0 \Rightarrow
W=-2U$, indicating hydrostatic equilibrium. Including a turbulent
velocity field, the condition of virial equilibrium becomes
$W=-2(U+K)$, indicating that turbulence can provide, in principle,
additional support against gravity. However, note that this assumes
that the turbulence is microscopic ({\it microturbulence}), since it
is implicitly required that the velocity field itself does not alter the mass
distribution within the volume.

\subsubsection{Magnetic support.}

Assume now a spherical cloud (a density peak) of radius $R$,
with a uniform field $\B = \Bo
\e_x$ inside and a vanishing field outside, and $\u = P =0$. It is
easy to show that for such a cloud, 
$W= -G \int [M(r)/r] dM(r) = -3GM^2/5R$ and $\M=\Bo^2 R^2/6 \equiv 
\phi^2/(6 \pi^2 R)$, where $M(r)$ is the mass interior to radius $r$ and
$\phi$ is the {\it magnetic flux}, which, according to the
flux-freezing condition, eq.\ (\ref{magnetic}), is conserved in the
absence of magnetic dissipation. Since $\phi$ is constant, 
both $W$ and $\M$ scale as $R^{-1}$ upon contraction, and the ratio
$W/\M$ is constant. This implies that there is a critical mass-to-flux
ratio $(M/\phi)_{\rm crit} \equiv (5/18)^{1/2} (\pi^2 G)^{-1/2}$
below which a cloud cannot collapse gravitationally, since the
magnetic energy $\M$ is larger than the gravitational energy $W$
(Strittmatter 1966). Modifications to this critical ratio due to the
geometry of the cloud were studied analytically also by Strittmatter
(1966) and numerically by Mouschovias \& Spitzer (1976), in the
presence of an external pressure. Bertoldi \& McKee (1992) have
studied the stability of 
spheroidal, pressure-confined clumps, giving estimates of the Jeans
and magnetic critical masses.

Finally, let us note that throughout this course, we discuss the {\it
scalar} version of the VT. A {\it tensor} form can also be derived,
which allows the treatment of multidirectional transport
phenomena. Zweibel (1990) has given a detailed account of magnetic
support and stability of spheroidal, quiescent cloud configurations
using the tensor VT.

\subsection{Properties of molecular clouds. Larson's
  relations}\label{larson relations}

Molecular clouds exhibit a number of correlations between their
various physical properties, such as their size, velocity dispersion,
magnetic field strength and (possibly) density and mass. Larson (1981)
first noticed, from a collection of data from several surveys
available to him, that the velocity dispersion $\sigma$ and number
density $n$ of molecular clouds scaled with their size as
\begin{equation}\label{origlars}
\Bigl({\sigma \over{\rm km\ s}^{-1}}\Bigr) \sim 1.1 \Bigl({R \over
  {\rm pc}}\Bigr)^{0.38},
\ \ \ \ \ \Bigl({n \over {\rm cm}^{-3}}\Bigr) \sim 3400 \Bigl({R
  \over{\rm pc}}\Bigr)^{-1.1}.
\end{equation}
A large number of subsequent studies (e.g., Torrelles et al.\ 1983;
Dame et al.\ 1986; Falgarone \& P\'erault 1987; Myers \& Goodman 1988a;
Falgarone et al.\ 1992; Miesch \& Bally 1994; Wood et al.\ 1994;
Caselli \& Myers 1995) have found similar scalings, although the
currently most favored forms are
\begin{equation}\label{larvel}
\sigma \sim R^{1/2}
\end{equation}
and
\begin{equation}\label{lardens}
\rho \sim R^{-1}
\end{equation}
where $\rho=mn$ is the mass density, and $m$ is the mean molecular mass.
These scaling relations have been interpreted, since their discovery,
as evidence for virial equilibrium of molecular clouds. Larson (1981)
noted that, as a consequence of these relations,
the ratio $2GM/\sigma^2 R$ was roughly constant for his cloud
sample, indicating virial balance between the turbulent velocity
dispersion and self-gravity ($W \sim K$). A problem with this result
however, was that the observed linewidths imply in general highly
supersonic velocity dispersions. Therefore, it was expected that such
random motions should produce shocks and dissipate rapidly, becoming
subsonic (e.g., Mestel 1965a, b).
This problem disappears, however, if one introduces magnetic fields
into the picture, since Alfv\'en waves do not 
dissipate as rapidly, and additionally
for clouds with nearly the critical mass-to-flux ratio, the Alfv\'en
velocity nearly equals the virial velocity (Shu et al.\ 1987; Mouschovias
1987). That is, from $W \sim {\cal M}$ one gets
\begin{equation}\label{magnetic bal}
{B^2 \over 4 \pi \rho} \equiv \va^2 \sim {GM \over R} \sim G\rho R^2, 
\end{equation}
while from $W \sim K$, we obtain
\begin{equation}\label{kinetic bal}
\sigma^2 \sim G\rho R^2,
\end{equation}
where $\va$ is the {\it Alfv\'en speed}.\footnote{Xie (1997) has pointed out
that it is actually the magnetic fluctuations $\Delta B$ that should
be considered in these expressions, rather than the mean field
$B$. However, the two generally turn out to be of comparable
magnitude (Myers \& Goodman 1991), 
and use of $B$ is justified, alhough the difference should
be kept in mind.} This in
principle could solve the dissipation problem, since in this case
velocity dispersions $\sigma$ 
up to the Alfv\'en speed do not induce shocks (although see
\S \ref{counterpoint}), while explaining the observed virial value of
the velocity dispersion. If additionally one assumes that for an
ensemble of clouds the density-size relation $\rho \sim R^{-1}$ 
is satisfied, then $\sigma^2 \sim G\rho R^3 /R \sim R$, automatically
satisfying the velocity dispersion-size relation. That is, out of the
two Larson's relations and the virial equilibrium condition, only two
are independent. Furthermore, the relation $\rho \sim R^{-1}$ can be
interpreted as a consequence of the clouds being near the critical
mass-to-flux ratio at roughly constant magnetic field
strengths. Indeed, from eq.\ (\ref{magnetic bal}) it follows that 
$\rho \sim R^{-1}$ for constant $B$.

\begin{figure}\label{MG88a1}
\psfig{figure=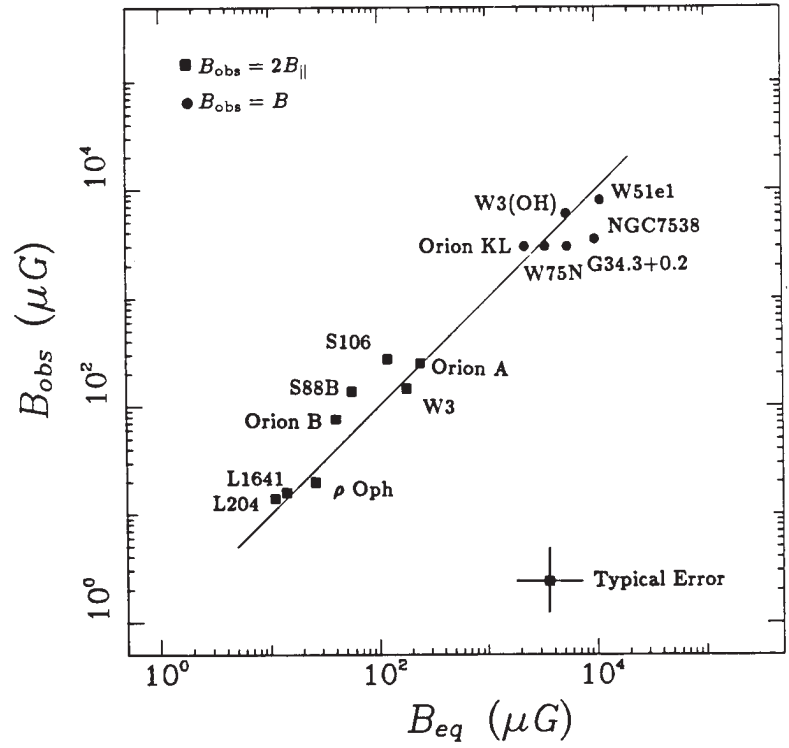}
\caption{Observed vs.\ virial equilibrium values of the magnetic field
  for the cloud sample of Myers \& Goodman 1988a.}
\end{figure}

Myers \& Goodman (1988a) tested these ideas on a cloud sample with
existing magnetic field strength determinations from Zeeman
measurements. Assuming $\sigma \sim \va$, it follows from eqs.\
(\ref{magnetic bal}) and (\ref{kinetic bal}) that the field strength
necessary for virial equilibrium is $B_{\rm eq} \sim \sigma^2/R$. 
Myers \& Goodman (1988a) plotted the observed
magnetic field strength versus $B_{\rm eq}$,
finding the two were equal to within factors of $\sim 3$ (fig.\ \ref{MG88a1}).

\begin{figure}\label{MG88a2}
\psfig{figure=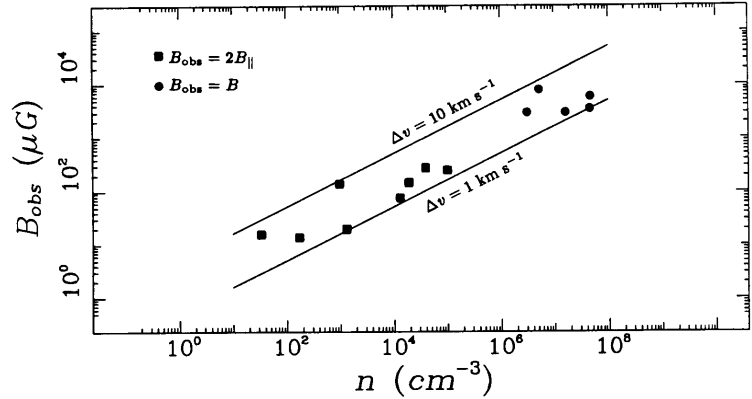,width=10cm}
\caption{Observed magnetic field strength vs.\ virial density $n\equiv 15
  (32 \pi \ln 2 m G)^{-1} (\Delta v/R)^2$ for the cloud sample of
  Myers \& Goodman 1988a. The solid lines indicate
  the virial equilibrium model for the indicated velocity dispersions.}
\end{figure}

Additionally, they also found 
that $B_{\rm obs} \sim \rho^{1/2}$, in agreement with
$\sigma^2 \sim B^2/\rho$ (again derived from combining relations
[\ref{magnetic bal}] and [\ref{kinetic bal}]), since in their
sample $\sigma$ had relatively little scatter compared with the
other quantities. The resulting
correlation had a scatter consistent with the scatter in $\sigma$
(fig.\ \ref{MG88a2}).
Note, however, that in the latter plot, the 
``virial'' density they used was an estimate
assuming virial equilibrium ($\rho \propto (\Delta v /R)^2$), 
not an actual observed value, and thus
this plot can only be taken as a proof of self-consistency of the
virial equilibrium hypothesis.

Myers \& Goodman (1988b) also considered the fact
that the observed linewidths contain both a thermal and a nonthermal
component, explaining deviations from Larson's (1981) trends at very
small scales (cloud cores of sizes $\sim 0.1$ pc). 
According to relation (\ref{larvel}), which refers to the macroscopic,
nonthermal velocity dispersion, at small enough cloud sizes the nonthermal
component becomes smaller than the thermal (the velocity dispersion
becomes subsonic). Indeed, Myers \& Goodman (1988b) find a drop in the
ratio of nonthermal to gravitational energy ($K_{\rm NT}/W$)
in their cloud sample for
small cloud sizes. It is important to note, 
however, that the trend is marginal when the raw data are used, 
but clearly noticeable only when the assumption
of virial equilibrium is introduced
(replacing the true gravitational energy obtained from direct
measurements of the size and density by the total kinetic (thermal +
nonthermal) energy). From this result, Myers \& Goodman (1988b) 
suggest that the uncertainties in $K_{\rm NT}/W$ arise more from
uncertainties in the density and size than from uncertainties in
velocity dispersion and tempertaure. However, an obvious alternative possibility
is that the scatter is real, indicating departures
from precise virial equilibrium. Their plot ``assuming virial
equilibrium'' is in reality just a plot of
the ratio of nonthermal to total kinetic energy, thus necessarily being
very tight at large scales if the velocity dispersion does 
increase with size
while the temperature remains roughly constant, but it is not 
a test of virial equilibrium in the clouds. On the other hand, further
supporting evidence for virial equilibrium has been provided by
Goodman et al.\ (1993), who find good agreement between the observed
and the ``virial'' values of $\beta$, the ratio of rotational to
gravitational energy in a sample of dense cores.

The role of thermal plus nonthermal (TNT) motions has been further
investigated by Fuller \& Myers (1992) and by Caselli \& Myers
(1995), finding respectively that the scaling of the nonthermal part
of the velocity differs from that of the total velocity dispersion,
and that the scaling is different for massive cores than for low-mass
cores. Furthermore, Fuller \& Myers (1992) have distinguished between {\it
cloud-to-cloud} scaling, i.e., that obtained when the sample
consists of various clouds observed with the same tracer (sensitive to
one particular density range) and {\it line-to-line} scaling, i.e.,
the scaling obtained when using various tracer molecules (sensitive to
a variety of density ranges) on a single cloud. 
The latter can be used to test for
virialization and the density structure of individual clouds as a
function of the distance from the core.
An even finer subdivision of possibilities has been proposed by
Goodman et al. (1997). These authors have additionally found that the
slope of the $\sigma$-$R$ relation using single-line, single-cloud
measurements, appears to decrease to nearly zero at scales $\sim 0.1$
pc, and that the column density $N$ changes its size dependence
from $N \sim R^{-0.2}$ at
scales $\simgt 0.1$ pc to $N \sim R^{-0.9}$ at scales $\simlt 0.1$ pc.
They interpret these results as evidence for a transition from non-thermal to
predominantly thermal support, and the onset of ``velocity coherence''
at those scales.

Finally, it should be noted that a number of other
possible mechanisms, capable of
originating Larson's relations or either one thereof, have been
proposed. For example, Larson (1981) himself suggested that the
density-size relation might be due to planar shocks which keep the
column density constant along the direction of shock
propagation. Chi\`eze (1987) has proposed that an ensemble of clouds 
on the verge of gravitational instability in an environment providing
external pressure should also satisfy Larson's relations. In the
chapter on Turbulence in Molecular Clouds, a discussion on mechanisms
related to hydro- or magnetohydrodynamic turbulence is given. Other
scenarios are unfortunately out of the scope of this course.

\subsection{Counterpoint}\label{counterpoint}

As it may have started to be apparent to the reader from the discussion of the
previous section, the issue of cloud virialization as the origin of
the Larson (1981) relations is not completely
settled. Besides the caveats inherent to the work of Myers \& Goodman
(1988a, b), which are the standard references for observational
determinations of virial equilibrium in clouds, other somewhat
contradictory evidence exists. 

First, let us distinguish between a cloud being in virial {\it
equilibrium} and it satisfying the VT. Actually, all clouds, and in
general all fluid parcels in the ISM, satisfy the VT, as it is a direct
consequence of the momentum conservation equation. Instead, virial
equilibrium means $d^2I/dt^2 =0$ specifically. 

\begin{figure}\label{MP95a}
\centerline{\psfig{figure=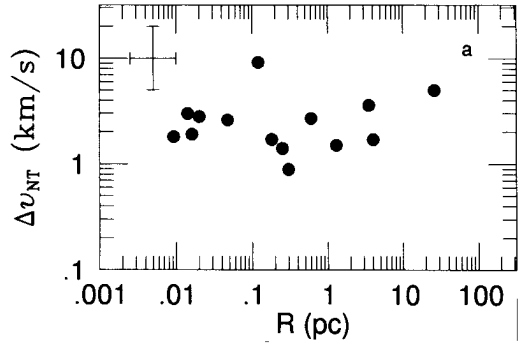}}
\caption{Nonthermal linewidth vs.\ size for 14 objects from Myers \&
  Goodman 1988a. No clear correlation is seen. (From Mouschovias \&
  Psaltis 1995.)}
\end{figure}
Concerning the velocity dispersion-size relation, Scalo (1990) and 
Mous\-chovias \& Psaltis (1995) have
pointed out that the original compiled by Myers \& Goodman (1988a) show
essentially just scatter for this relation (fig.\ \ref{MP95a}). 
Mouschovias \& Psaltis (1995) ``solve the problem'' by
plotting instead the ratio $\sigma/R^{1/2}$ vs.\ the
magnetic field strength $B$. In reality, however, this is nothing but
the same $B_{\rm obs}$ vs.\ $B_{\rm eq}$ plot shown by Myers \&
Goodman (1988a). Moreover,
Mouschovias \& Psaltis (1995) use this result as
an argument in favor of the $\sigma$-$R$ relation being a consequence
of magnetic support. However, they arrive at this conclusion by
assuming $B$ does not ``vary much from place to place in the ISM under
conditions suitable for the formation of self-gravitating clouds''; yet,
the magnetic field strength data they consider vary by three orders of
magnitude! In summary, the cloud data discussed in Myers \&
Goodman (1988a) are consistent with magnetic
support, but do not satisfy the dispersion-size relation.

%

Furthermore, there exists both observational and
numerical evidence that the density-size relation is not always
satisfied, implying that the column density is not constant. It was
already pointed out by Larson (1981) himself that the density-size
relation might not be a real property of molecular clouds, but instead
just an artifact of limitations of the observational methods used to
produce cloud surveys. Kegel (1989) and Scalo (1990) elaborated on
this issue. Also, a number of observational studies have found significantly
different scaling exponents for both Larson's relations, or no clear
scaling at all. For example, Carr (1987) has found $\sigma \sim
R^{0.25}$ and $M \sim R^{2.5}$, implying $\rho \sim R^{-0.5}$ for 45
clumps in the Cep OB 3 molecular cloud. In
fact, Carr notes that the clumps are not in virial equilibrium, and
concludes that they must be expanding. Loren (1989) studied 89 clumps
in the $\rho$ Oph molecular cloud, finding $\rho \sim R^{-0.2}$ and no
$\sigma$-$R$ correlation. He points out that many of the clumps are
extremely filamentary and that they may probably be the outcome of the
passage of a shock wave. Falgarone et al.\ (1992) focused on low
brightness areas of several molecular clouds, including both
gravitationally bound and unbound structures (structures for which the
actual masses are much smaller than the value obtained assuming virial
equilibrium), finding $\sigma \sim R^{0.4}$, and a density-size plot with
large scatter and an {\it upper bound} given by a Larson-type relation
$\rho \sim R^{-1}$. V\'azquez-Semadeni et al.\ (1997) have found the
same effect in numerical simulations
of turbulent cloud formation in the ISM, in which in fact there is no
unique $\rho$-$R$ correlation. Plume et al.\ (1997), in a
sample of maser-selected, massive star formation regions, find no
significant $\sigma$-$R$ or $\rho$-$R$ correlations either.

Finally, Myers et al.\ (1995) have found an example of a cloud in
which the kinetic and magnetic energies are comparable, yet the
gravitational energy is roughly 500 times smaller, indicating that
self-gravity is negligible for this cloud. 

\section{Discussion and conclusions}\label{conclusions}

From the discussion in the previous sections, a number of points can be
concluded.

\medskip
\noindent
1. ``Virial equilibrium'',
i.e., $d^2I/dt^2=0$ allows up to a constant rate of change in the
moment of inertia $I$, even though it is commonly taken to mean a
stationary equilibrium.

\medskip
\noindent
2. Larson's (1981) relations, eqs.\ (\ref{larvel}) and
(\ref{lardens}), have been interpreted as a consequence of virial
equilibrium\footnote{Mouschovias \& Psaltis (1995) distinguish between
  magnetic virial equilibrium and magnetic support, but the resulting
  balance equation is the same.} between self-gravity on the one hand 
and turbulent velocity dispersion and/or magnetic support on the
other. From the observational data discussed above,
this interpretation may apply to strongly condensed
clumps which, due to self-gravity, have decoupled from their
surrounding medium, and thus require internal support against collapse.

\medskip
\noindent
3. The above interpretation, however, is insufficient for explaining
{\it both} scaling relations. One is still independent and requires an
additional explanation. A critical mass-to-flux ratio has been
invoked as the origin of the density-size relation assuming roughly
constant magnetic field strengths (Shu et al.\ 1987; Myers \& Goodman
1988b; Mouschovias \& Psaltis 1995), but this assumption is not verified
in the observational data. For example, the magnetic field strenghts
discussed by Myers \& Goodman (1988a) and Mouschovias and Pslatis
(1995) span three orders of magnitude. 
Besides, examples of virialized clouds  which
however follow different scaling relations exist (e.g., Loren 1989;
Fuller \& Myers 1992; Caselli \& Myers 1995; Goodman et al.\ 1997; see also
V\'azquez-Semadeni \& Gazol 1995 for a theoretical discussion). Planar
shocks have also been invoked as an explanation of the apparently
constant column densities (Larson 1981; Scalo 1987).

\medskip
\noindent
4. It is possible that the density-size relation is not a true
property of clouds, but only an observational effect due to the
limited integration times used in surveys (Larson 1981; Kegel 1989;
Scalo 1990), which tend to select constant-column density obejcts. 
In this case, surveys using larger integration times would
exhibit an increasingly larger scatter in density-size
plots, as in the data of Falgarone et al.\ (1992) or the numerical
data of V\'azquez-Semadeni et al.\ (1997), the latter being free from such
limitations. In fact, large scatter was already present in the Myers \&
Goodman (1988b) data. Although they interpreted it as uncertainty in the
data rather than real scatter about virialization, the latter
possibility is equally feasible. A study that seemed to find
exceptionally constant column densities while claiming a large dynamic
range (Wood et al.\ 1994) has been questioned by \VS\ et al.\ (1997).

\medskip
\noindent
5. The velocity dispersion-size relation does not appear subject to the
possibility of a spurious origin, and therefore it is probably real
{\it when detected}, although in many cases it is not (e.g., Loren
1989; Plume et al.\ 1997). For the cases when it is present, a number
of physical mechanisms are plausible candidates. If the density-size
relation is not verified, then the standard arguments based on virial
equilibrium between gravity and velocity dispersion
cannot be invoked (recall the $\sigma$-$R$ relation is a
consequence of virial equilibrium {\it and} the density-size
relation). In particular, the possibility that the $\sigma$-$R$ relation is a
consequence of the characteristic 
energy spectrum of an ensemble of shocks in the
flow has been suggested by a number of authors (e.g., Passot et al.\
1988; Padoan 1995; Gammie \& Ostriker 1996; Fleck 1996; \VS\ et al.\
1997). In this case, the origin of the dispersion-size relation would
be completely independent of the virial condition, explaining why the relation
can be present even in the absence of a density-size relation.

\medskip
\noindent
6. In summary, the Larson's relations and their virial equilibrium
interpretation probably apply to relaxed, strongly self-gravitating clumps.
Regions which do not exhibit clear scaling relations often include
either massive cores (Caselli \& Myers 1995; Plume et al.\ 1997) 
or regions with strong evidence of recent perturbations (Loren
1989). In fact, the dense cores studied by Plume et al.\ were selected
by the presence of H$_2$O masers, suggesting strong excitation
mechanisms as well. Such ``perturbed'' regions are likely not in
virial equilibrium, thus also being transients (e.g., Magnani et
al.\ 1993), or at least having
strongly fluctuating moments of inertia (shapes and internal mass
distributions).

\vspace*{0.3cm}
\acknowledgements 

The author wishes to thank A.\ Goodman, T.\ Passot and J.\ Scalo for a critical
reading of the manuscript, as well as helpful comments and precisions.


\end{document}